\documentclass{article}
\usepackage{amsmath,amssymb,theorem,cite}
\theorembodyfont{\upshape}
\newtheorem{theorem}{Theorem}[section]   
\newtheorem{lemma}[theorem]{Lemma}
\numberwithin{equation}{section}
\numberwithin{theorem}{section}

\newcommand{\due}[2]{\genfrac{}{}{0pt}{1}{#1}{#2}}   
\newcommand{\qed}{\hfill $\blacksquare$} 
\newcommand{\Z}{\mathbb Z}       
\newcommand{\R}{\mathbb R}
\newcommand{\N}{\mathbb N}       

\begin{document}

\title{On the propagation of a perturbation in an 
anharmonic system}

\author{Paolo Butt\`a\footnote{Dipartimento di Matematica, 
Universit\`a di Roma `La Sapienza', P.le Aldo Moro 2, \hfill\break
00185 Roma, Italy. E-mail: {\tt butta@mat.uniroma1.it,
caglioti@mat.uniroma1.it, diruzza@mat.uniroma1.it,
marchior@mat.uniroma1.it.} Fax: +39--06--44701007.}
\and Emanuele Caglioti$^*$ \and Sara Di Ruzza$^*$ 
\and Carlo Marchioro$^*$}

\date{}
\maketitle

\begin{abstract} 
We give a not trivial upper bound on the velocity of disturbances 
in an infinitely extended anharmonic system at thermal equilibrium.
The proof is achieved by combining a control on the non equilibrium
dynamics with an explicit use of the state invariance with respect to
the time evolution.
\end{abstract}


\medskip\noindent
{\bf Running title:} Propagation of a perturbation in an anharmonic
system.

\section{Introduction}
\label{sec:1}

In the present paper we investigate the long time behavior of an
infinitely extended anharmonic system, that represents, of course, a
schematic model of a crystal. There are several papers devoted to 
the time evolution of systems containing an infinite number of
components, either particles moving in a continuum or lattice 
systems, see Refs.\ 
\cite{BPY,BCCM,CMP,CC,CMS,DF,F,FD,LanI,LanII,LLL,MPP1,MPP2,MP}, 
but few of them study the asymptotic (in time) behavior of the system. 
The reason lies in the difficulty to obtain dynamical estimates that 
remain good for very long times. Some results in this direction have
been recently obtained \cite{BCM1,BCM2,BMM,CM}, but they are not related
to the topic of the present paper.

Such dynamical estimates can be greatly improved if the physical
system is assumed at thermal equilibrium. Actually, it is not 
necessary to consider exactly a Gibbs state: the present results apply
to any reasonable time invariant (i.e.\ stationary) state. 
This assumption allows to prove the locality in space of the motion
and hence its existence, see Refs.\ \cite{A,LanIII,MPPr,PPT,P,SS,S1,S2}.
Moreover, in this case it is known (but perhaps not sufficiently
underlined) that the magnitude of a component of the system may  
increase only very slowly in time. This fact allows us to obtain not 
trivial results for the following problem. 

Consider a lattice system in dimension $d$, with anharmonic oscillators 
at each site. At time zero we perturb the oscillator
located on the site $i$ and study the influence at time $t$ on that
located on the site $j$. For harmonic oscillators or bounded rotators
it has been proved in Ref.\ \cite{MPPT} that this influence becomes
negligible when $|i-j|>c t$, for some $c$ which thus gives a bound 
of the velocity of propagation of the perturbation. 

The case of anharmonic oscillators is much more involved. The general 
case seems too difficult, while for systems in thermal equilibrium 
a reasonable estimate has been obtained in Ref.\ \cite{MPPT}: the
influence of the perturbation is exponentially small in time for 
distances $|i-j| > t^{4/3}$.

At this point we must precise what we mean by perturbation. 
Actually, we do not modify the equilibrium state, rather we analyze  
observables which give an estimate of the size of the correlation. 
Following Ref.\ \cite{MPPT}, we use the Poisson brackets to this 
purpose, by analogy with a similar problem in the quantum case  
previously studied in Ref.\ \cite{LR}. More precisely, we analyze the
Poisson brackets $\{f_i,g_j\circ \Phi_t \}$ where $f_i,g_j$ are
observables localized on the site $i,j$ respectively, and $\Phi_t$
denotes the time evolution. We prove that,
with probability one with respect to any reasonable stationary state,
$\{f_i,g_j\circ \Phi_t \}$ is exponentially small in time whenever 
$|i-j| > t\, \log^\alpha t$ for a suitable $\alpha>0$. 
This gives a non trivial bound on the velocity of the perturbation. 

The problems and the results are rigorously posed in Section 
\ref{sec:2} and proved in Section \ref{sec:3}. 

\section{Notation and statement of the result}
\label{sec:2}

At each point $i$ of the $d$-dimensional lattice $\Z^d$ there is an
oscillator with coordinate $q_i\in\R$ and momentum $p_i\in\R$. The
state of the system is thus determined by the infinite sequence $x =
\{x_i\}_{i\in\Z^d} = \{(q_i,p_i)\}_{i\in\Z^d}$ of positions and
momenta of the oscillators.  We shall denote by $\mathcal{X}$ the set
of all such possible states, equipped with the product topology.

The time evolution $t\mapsto x(t) = \{(q_i(t),p_i(t))\}_{i\in\Z^d}$ 
is defined by the solutions of the following infinite set of coupled
differential equations,
\begin{equation}
\label{p1}
\begin{cases} \dot q_i (t) = p_i(t) \\
\dot p_i(t) = F_i(x(t)) \end{cases} \qquad i \in \Z^d,
\end{equation}
where the force $F_i(x)$ induced by the configuration
$x=\{(q_i,p_i)\}_{i\in\Z^d}$ on the $i$-th oscillator is given by
\begin{equation}
\label{p2}
F_i(x) = - U'(q_i) - K \sum_{j:|j-i|=1} (q_i-q_j),
\end{equation}
with $K>0$ and $U(\cdot)$ a non negative polynomial of degree $4$ with 
strictly positive leading coefficient. Here $|i-j|$ is the distance
between the points $i=(i_1,\ldots,i_d)$ and $j=(j_1,\ldots,j_d)$ defined by
\begin{equation*}
|i-j| = \sum_{\ell=1}^d |i_\ell-j_\ell|.
\end{equation*}

In order to consider configurations that are typical for any
reasonable thermodynamic equilibrium state, we allow initial data with          
logarithmic divergences in the energy. More precisely, for
$\nu\in\Z^d$ and $k\in\N$, let
\begin{equation}
\label{p3}
W_{\nu,k}(x)  \doteq \sum_{i\in\Lambda_{\nu,k}} \bigg\{ 
\frac{p_i^2}{2} + U(q_i) + 1 \bigg\} 
+ \sum_{\due{i,j\in\Lambda_{\nu,k}}{|j-i|=1}} \frac K4 (q_i-q_j)^2,
\end{equation}
where $\Lambda_{\nu,k}$ denotes the cube of center $\nu$ and side
$2k+1$. By defining 
\begin{equation}
\label{p4}
Q(x) \doteq \sup_{\nu\in\Z^d} \,\, \sup_{k>\log^{1/d}(e+|\nu|)} \,\,
\frac{W_{\nu,k}(x)}{(2k+1)^d},
\end{equation}
we denote by $\mathcal{X}_0$ the following subset of $\mathcal{X}$,
$$
\mathcal{X}_0 = \{ x\in \mathcal{X} \, : \, Q(x)<\infty\}.
$$ 

Let now $\omega$ be any Borel probability measure on $\mathcal{X}$ that 
satisfies the following superstability estimate: there exists a
positive constant $C_\omega$ such that, for any $\lambda$ small enough,
\begin{equation}
\label{p5}
\omega\big( e^{\lambda W_{\nu,k}}\big) \leqslant e^{C_\omega (2k+1)^d}
\qquad \forall\, \nu\in\Z^d \quad \forall\, k\in \N.
\end{equation}
It is proved in the Appendix that this implies, for any 
$\lambda$ small enough,
\begin{equation}
\label{p6}
\lim_{N\to + \infty} e^{\lambda N} \, \omega(Q>N) = 0,
\end{equation}
which in particular yields $\omega(\mathcal{X}_0) = 1$. 

The following theorem gives existence and uniqueness of the solution
to Eq.\ \eqref{p1} for initial data in the set $\mathcal{X}_0$.

\begin{theorem}
\label{t:p1}
There exists a one-parameter group of transformations $\Phi_t :
\mathcal{X}_0 \to \mathcal{X}_0$, $t\in \R$, such that $t \to 
\Phi_t(x)$ is the unique global solution to Eq.\ \eqref{p1} with initial 
condition $\Phi_0(x)=x$. Moreover $\Phi_t(x)_j = x_j(t) 
= (q_j(t),p_j(t))$ are differentiable functions with respect to 
$x_i = (q_i,p_i)$ for any $j,i\in\Z^d$ and $t\in\R$. Finally, there 
exists a constant $C_0$ such that
\begin{equation}
\label{p7}
Q(\Phi_t(x)) \leqslant C_0 \Big\{ Q(x) \log [e+Q(x)] + t^4 \Big\}.
\end{equation}
\end{theorem}

The proof of Theorem \ref{t:p1} is given in the Appendix. We remark
that a global existence and uniqueness theorem for a larger class of initial 
data can be found in Ref.\ \cite{LLL}. Unfortunately, the proof
given there does not guarantee that the set $\mathcal{X}_0$ is invariant
under the dynamics. We instead adapt to the present context the
technique developed by Dobrushin and Fritz in the case of particles
in the continuum \cite{FD,DF}.

Let $\mathcal{U}$ be the algebra of all local observables. Thus
$f:\mathcal{X}\to \R$ is an element of $\mathcal{U}$ iff $f(x) =
f_\Lambda(x_\Lambda)$ for some bounded set $\Lambda\subset\Z^d$ and
some differentiable function $f_\Lambda$, depending on the finite set 
of real variables $x_\Lambda = \{x_i\}_{i\in\Lambda}$, which is 
bounded with its derivatives. The action of the time evolution on the
local observables is still denoted by $\Phi_t$: if $f\in\mathcal{U}$ the 
function $\Phi_tf$ is defined by setting $\Phi_tf(x) = f(\Phi_t(x))$
for any $x\in\mathcal{X}_0$. We finally observe that the Poisson
brackets 
$$
\{f,g\}(x) = \sum_{i\in\Z^d} \bigg(\frac{\partial f}{\partial q_i}
\frac{\partial g}{\partial p_i} - \frac{\partial f}{\partial p_i}
\frac{\partial g}{\partial q_i}\bigg)(x)
$$
are well defined for any $f,g\in \mathcal{U}$. 
                                                                                
A time invariant state $\omega$ is a probability measure on 
$\mathcal{X}$ for which \eqref{p5} holds and such that 
$\omega(\Phi_t f) = \omega(f)$ for any $f\in\mathcal{U}$ and $t\in\R$. 
The typical example of such a state is given by the infinite Gibbs
measure obtained as the thermodynamic limit with free boundary
conditions \cite{R}. 

Given a pair of differentiable function $f,g:\R^2\to \R$, which are
bounded with their derivatives, we denote by $f_i$, resp.\ $g_j$, the
local observable defined by setting $f_i(x) = f(x_i)$, resp.\
$g_j(x) = g(x_j)$. Our main result concerns the asymptotic behavior 
of the Poisson brackets $\{f_i,\Phi_t g_j\}(x)$ in the limit when
simultaneously $|i-j|\to\infty$ and $t\to\infty$.

\begin{theorem}
\label{t:p2}
Let $\omega$ be any time invariant state satisfying \eqref{p5}. Then 
for each $f,g$ as above, $\alpha > 1/2$, and $b>0$ we have
\begin{equation}
\label{p8}
\lim_{t\to\infty} \sup_{j\,:\,|i-j|> t \log^\alpha t}
e^{bt} \, \{f_i,\Phi_t g_j\}(x)  = 0,
\end{equation}
almost surely with respect to the probability measure $\omega$.
\end{theorem}

\medskip\noindent
{\bf Remark.} We consider the case of one dimensional oscillators 
with quartic one-body potential and nearest-neighbor harmonic
interaction only for the sake of simplicity. Indeed, the same strategy
applies to the general case of $N$-dimensional oscillators, say 
$q_i\in\R^N$, such that the force $F_i(x)$ induced by the configuration 
$x=\{(q_i,p_i)\}_{i\in\Z^d}$ on the $i$-th oscillator is now given by
\begin{equation*}
F_i(x) = - \nabla_{q_i} U(q_i) - \sum_{j:|j-i|\leqslant r} 
\nabla_{q_i} V(q_i-q_j). 
\end{equation*}
Here $r$ is a positive parameter, while $U$ and $V$ are smooth functions
such that $U(\xi) = P_1(|\xi|)$ and $V(\xi)= P_2(|\xi|)$, where $P_1$
and $P_2$ are non negative polynomials with maximum degree respectively
$2\gamma$ and $\leqslant \gamma$. The estimate \eqref{p8} is now
valid for any $\alpha>(\gamma-1)/\gamma$.

\bigskip
We conclude the section with a notation warning: in the sequel, if not
further specified, we shall denote by $C$ a generic positive constant
whose numerical value may change from line to line and it may possibly
depend only on the coupling constant $K$ and the one body interaction
$U(\cdot)$.

\section{Proof of Theorem \ref{t:p2}}
\label{sec:3}

In this section we prove Theorem \ref{t:p2}. We remark that the time 
invariance and the assumption \eqref{p5} are the only hypothesis
about the state $\omega$ used in the proof. Hence, even if $\omega$ is
not a spatially homogeneous state, we can assume $i=0$ without loss 
of generality (otherwise stated, the rate of convergence in \eqref{p8}
turns out to be estimated uniformly with respect to the location of
the site $i$). 

Let us briefly outline the strategy of the proof. In the case of
anharmonic systems, the unboundedness of the Lipschitz constant of the
force is the source of troubles in obtaining not trivial dynamical
estimates. To overcome this problem, we consider a ``good set'' of
initial data, the set $\mathcal{B}$ defined below, where the dynamics
is under control for all sufficiently large times $k\in\N$. By 
\eqref{p6} and the time invariance of $\omega$, it is readily seen
that $\omega(\mathcal{B})=1$.
On the other hand, using rough a priori bounds on the dynamics for 
short times, an obvious interpolation shows that in the set 
$\mathcal{B}$ the time evolution $\Phi_t$ is under control for any 
$t\geqslant 0$. Such a control turns out to be good enough to solve and
bound, by iteration, the (linear) variational equation for the disturbance. We now proceed with the proof.

\bigskip
Given $\alpha>1/2$ as in the statement of Theorem \ref{t:p2}, we
choose $\delta \in (1, 4\alpha-1)$ and define
$$
\mathcal{B} \doteq \bigcup_{n=1}^\infty \bigcap_{k=n}^\infty 
\mathcal{B}_k, \qquad \mathcal{B}_k \doteq \big\{x\in\mathcal{X} 
\, :  \,\, Q(\Phi_k(x)) \leqslant \log^\delta k \big\}.
$$
Setting $\mathcal{B}_k^\complement \doteq \mathcal{X} \setminus 
\mathcal{B}_k$, since $\omega$ is time invariant, we have
$$
\omega\big(\mathcal{B}_k^\complement \big) = 
\omega\big(Q \circ \Phi_k > \log^\delta k \big) =
\omega \big(Q > \log^\delta k \big).
$$
It follows, by \eqref{p6}, that if $\lambda$ is small enough then 
$e^{\lambda\log^\delta k}\omega\big(\mathcal{B}_k^\complement 
\big) \to 0$ as $k \to \infty$. In particular, since $\delta>1$, 
$\sum_k\omega \big(\mathcal{B}_k^\complement \big) < \infty$, whence 
\begin{equation}
\label{b=1}
\omega(\mathcal{B})=1
\end{equation}
by the Borel-Cantelli lemma.
 
We next estimate:
\begin{eqnarray}
\label{p9}  
\big|\{f_0,\Phi_t g_j\}\big|(x) & = & \bigg|\frac{\partial f_0}{\partial
q_0} \frac{\partial (\Phi_t g_j)}{\partial p_0} - \frac{\partial f_0}
{\partial p_0} \frac{\partial (\Phi_t g_j)}{\partial q_0}\bigg|(x)
\nonumber \\ \nonumber 
&  = & \bigg|\frac{\partial f_0}{\partial q_0} 
\bigg(\frac{\partial g_j}{\partial q_j}\bigg|_{\Phi_t(x)} 
\frac{\partial q_j(t)}{\partial p_0} + 
\frac{\partial g_j}{\partial p_j}\bigg|_{\Phi_t(x)}
\frac{\partial p_j(t)}{\partial p_0} \bigg)
\\ \nonumber && -\, \frac{\partial f_0}{\partial p_0} 
\bigg(\frac{\partial g_j}{\partial q_j}\bigg|_{\Phi_t(x)} 
\frac{\partial q_j(t)}{\partial q_0} + 
\frac{\partial g_j}{\partial p_j}\bigg|_{\Phi_t(x)}
\frac{\partial p_j(t)}{\partial q_0} \bigg)\bigg|(x)
\\ & \leqslant & 4 \, \|\nabla f\|_\infty \, \|\nabla g\|_\infty 
\, \big\|\Delta_j(t,x)\big\|.
\end{eqnarray}
Here $(q_j(t),p_j(t))$ denote the $j$-th coordinates of $\Phi_t(x)$
and $\big\|\Delta_j(t,x)\big\|$ is the uniform norm of the 
$2\times 2$ Jacobian matrix $\Delta_j(t,x)$ given by 
\begin{equation}
\label{p10}
\Delta_j(t,x) = \frac{\partial \Phi_t(x)_j}{\partial x_0} =
\left(\begin{array}{cc}
{\displaystyle \frac{\partial q_j(t)}{\partial q_0}} &  
{\displaystyle \frac{\partial q_j(t)}{\partial p_0}} \\ \\
{\displaystyle \frac{\partial p_j(t)}{\partial q_0}} &
{\displaystyle \frac{\partial p_j(t)}{\partial p_0}} 
\end{array} \right).
\end{equation}
Hence, by \eqref{b=1} and \eqref{p9}, Theorem \ref{t:p2} follows once
we show that, for any $b>0$,
\begin{equation}
\label{p11}
\lim_{t\to\infty} \sup_{j\,:\,|j|> t \log^\alpha t} e^{bt}\, 
\big\| \Delta_j(t,x)\big\| = 0 \qquad \forall\, x\in\mathcal{B}. 
\end{equation}

\begin{lemma}
\label{t:p3}
There exists a positive constant $C_1>0$ such that, for any $x\in 
\mathcal{X}_0$, $j\in\Z^d$, and $t > 0$,
\begin{equation}
\label{p12}
\big\|\Delta_j(t,x)\big\| \leqslant (1+t) \sum_{n= |j|}^\infty 
\bigg(\frac{H(t,x)\log^{1/2}(e+n)}{n^2}\bigg)^n,
\end{equation}
where
\begin{equation}
\label{p13}
H(t,x) = C_1 \, t^2 \, \Big[\sup_{0\leqslant s\leqslant t} 
Q(\Phi_s(x))\Big]^{1/2}.
\end{equation}
\end{lemma}

\medskip\noindent
{\bf Proof.}
By \eqref{p1}, the trajectory $(q_j(t),p_j(t)) = \Phi_t(x)_j$ satisfies 
the equation
\begin{equation*}
\begin{pmatrix} q_j(t) \\ p_j(t) \end{pmatrix} =
\begin{pmatrix} q_j + p_j t \\ p_j \end{pmatrix} +
\int_0^t\!ds\, \begin{pmatrix} (t-s) F_j(\Phi_s(x)) \\ F_j(\Phi_s(x))
\end{pmatrix}, 
\end{equation*}
from which we get, recalling \eqref{p2},
\begin{equation}
\label{p14}
\Delta_j(t,x) = \begin{pmatrix}1 & t \\ 0 & 1 \end{pmatrix} 
\delta_{0,j} + \sum_{h\in\Z^d} \,  \int_0^t\!ds\, B_{j,h}(s) 
\begin{pmatrix}t-s & 0 \\ 1 & 0 \end{pmatrix} \Delta_h(s,x), 
\end{equation}
with
\begin{equation}
\label{p15}
B_{j,h}(s) = - \big[U''(q_j(s))+2dK\big] \, \delta_{j,h} +
K \sum_{\ell : |\ell-j|=1} \delta_{\ell, h}.
\end{equation}
The integral equation \eqref{p14} can be solved by iteration,
getting
\begin{eqnarray}
\label{p16}
\Delta_j(t,x) & = & \begin{pmatrix}1 & t \\ 0 & 1 \end{pmatrix} 
\delta_{0,j} + \sum_{n=1}^\infty \int_0^t\!dt_1
\int_0^{t_1}\!dt_2 \cdots \int_0^{t_{n-1}}\!dt_n\,
G_j(t_1,\ldots,t_n) 
\nonumber \\ && \times \, 
\begin{pmatrix}(t-t_1)(t_1-t_2)\cdots(t_{n-1}-t_n) & 0 \\ (t_1-t_2)\cdots(t_{n-1}-t_n) & 0 \end{pmatrix}
\begin{pmatrix}1 & t_n \\ 0 & 1 \end{pmatrix},
\end{eqnarray}
where
\begin{equation*}
G_j(t_1,\ldots,t_n) = \begin{cases}
{\displaystyle \sum_{k_1,\ldots,k_n}} B_{0,k_1}(t_1) \cdots
B_{k_n,j}(t_n) & \text{ if }\, |j|\leqslant n, \\ 0 & \text{ otherwise}.
\end{cases}
\end{equation*}
From the definition \eqref{p15}, by using \eqref{p3}, \eqref{p4}, 
and recalling $U(\cdot)$ is assumed a non negative polynomial of 
degree $4$ with strictly positive leading coefficient, it follows 
that
\begin{equation}
\label{p17}
\big| B_{k,h}(s) \big| \leqslant C \sqrt{Q(\Phi_s(x)) \log(e+|k|)}
\qquad \forall\, k,h\in\Z^d \quad \forall\, s\in\R. 
\end{equation}
We now observe that the sum in \eqref{p16} involves only sites $k_\ell$ 
such that $|k_\ell| \leqslant n$. On the other hand, the number of 
$n$-step walks, starting from a given site and with the possible 
presence of some permanences, is bounded by
$(2d+1)^n$. Then, by \eqref{p16} and \eqref{p17},
\begin{equation*}
\big\|\Delta_j(t,x)\big\| \leqslant  (1+t) \sum_{n = |j|}^\infty 
\frac{\big[C\,t^2\, \log^{1/2}(e+n)\big]^n}{(2n)!}
\Big[\sup_{0\leqslant s\leqslant t} Q(\Phi_s(x))\Big]^{n/2}. 
\end{equation*}
By the Stirling formula the bound \eqref{p12} now follows for a suitable 
choice of $C_1$ in \eqref{p13}.
\qed

\bigskip
We can now prove \eqref{p11}. Let $x\in \mathcal{B}$, by the definition
of $\mathcal{B}$ there exists a positive integer $n_0 = n_0(x)$ such
that $Q(\Phi_k(x)) \leqslant \log^\delta k$ for any $k\geqslant n_0$. 
By \eqref{p7}, for any $t>n_0$, 
\begin{eqnarray*}
\sup_{0\leqslant s\leqslant t} Q(\Phi_s(x)) & \leqslant  & 
\sup_{0\leqslant s \leqslant n_0} Q(\Phi_s(x)) +
\max_{k=n_0,\ldots, [t]}\: \sup_{0\leqslant s\leqslant 1} 
Q\big(\Phi_s\big(\Phi_{k}(x)\big)\big)
\\& \leqslant & C_0 \big\{ Q(x) \log [e+Q(x)] + n_0^4 \big\}
\\&& +\, \max_{k=n_0,\ldots, [t]}\:
C_0 \big\{ Q\big(\Phi_k(x)\big) \log [e+Q(\Phi_k(x))] + 1 \big\},
\end{eqnarray*}
whence, for $t_0=t_0(x)$ sufficiently large,
$$
\sup_{0\leqslant s\leqslant t} Q(\Phi_s(x)) \leqslant C 
\log^\delta t \log\log t \qquad \forall\, t\geqslant t_0.
$$
By \eqref{p12} and \eqref{p13} we thus obtain, for any $t$ 
large enough,
$$
\big\|\Delta_j(t,x)\big\|  \leqslant (1+t) \sum_{n= |j|}^\infty 
\left(\frac{C t^2 \sqrt{\log(e+n)\log^\delta t
\log \log t}}{n^2}\right)^n.
$$
If $|j|>t\log^\alpha t$, recalling $\delta<4\alpha-1$, the right 
hand side is readily seen to be bounded by $\exp\big\{-t\log^{1/2}
t\big\}$ for $t$ sufficiently large. The limit \eqref{p11} is thus proved.

\section*{Appendix}

\setcounter{equation}{0}
\def\theequation{A.\arabic{equation}}

In this appendix we prove Eq.\ \eqref{p6} and Theorem \ref{t:p1}.

\bigskip 
\noindent
{\bf Proof of Eq.\ (\ref{p6}).}
From the definition \eqref{p4} we have
\begin{equation*}
\omega(Q>N) \leqslant \sum_{\nu\in\Z^d} \, \sum_{k>\log^{1/d}(e+|\nu|)} 
\omega\big( W_{\nu,k}> N (2k+1)^d\big).
\end{equation*}
Applying the Exponential Chebyshev Inequality and the assumption 
\eqref{p5} on the measure $\omega$, we next bound, for some 
$\lambda_0>0$ sufficiently small,
\begin{equation*}
\omega\big( W_{\nu,k}> N (2k+1)^d\big) \leqslant \omega\big(
e^{\lambda_0( W_{\nu,k} - N (2k+1)^d)}\big) \leqslant
e^{-(N\lambda_0 - C_\omega)(2k+1)^d}. 
\end{equation*}
Then, for any $N >(d+2C_\omega)/\lambda_0$,
\begin{eqnarray*}
\omega(Q>N) & \leqslant & \sum_{\nu\in\Z^d} \, 
\sum_{k>\log^{1/d}(e+|\nu|)}e^{-N\lambda_0(2k+1)^d/2} 
\\ & \leqslant & C \sum_{\nu\in\Z^d}\frac{1}{(e+|\nu|)^{N\lambda_0}}
\, \leqslant \, C e^{-N\lambda_0},
\end{eqnarray*}
which implies \eqref{p6} for any $\lambda<\lambda_0$. 
\qed

\bigskip
\noindent
{\bf Proof of Theorem \ref{t:p1}.}
For any initial condition $x=\{x_i\}_{i\in \Z^d}\in \mathcal{X}_0$,
the solution to Eq.\ \eqref{p1} is constructed as the limit
\begin{equation}
\label{ap1}
\Phi_t(x)_i = \lim_{n\to \infty} \Phi_t^{(n)}(x)_i \qquad 
\forall\, i\in\Z^d,
\end{equation}
where the {\it $n$-partial dynamics} $\Phi_t^{(n)}(x)$ is defined
in the following way. For any $n\in\N$ let $\Lambda_n=\Lambda_{0,n}$
be the cube of side $2n+1$ centered in the origin. Then
$\Phi_t^{(n)}(x) = \big\{(q^{(n)}_i (t), p^{(n)}_i(t))
\big\}_{i\in\Lambda_n}$ is the solution to the Cauchy problem
\begin{equation}
\label{ap2}
\begin{cases} \dot q^{(n)}_i (t) = p^{(n)}_i(t), \\ 
\dot p^{(n)}_i(t) = F_i(\Phi_t^{(n)}(x)), \\
(q^{(n)}_i (0), p^{(n)}_i(0)) = x_i, \quad\qquad\qquad 
i\in \Lambda_n, \end{cases} 
\end{equation}
where
\begin{equation}
\label{ap3}
F_i(\Phi_t^{(n)}(x)) =  - U'(q^{(n)}_i(t)) - K
\sum_{j\in\Lambda_n :|j-i|=1} \big[q^{(n)}_i(t)-q^{(n)}_j(t)\big].
\end{equation}

\begin{lemma}
\label{l:ap1}
There exists a positive constant $C_2$ such that, for any $x\in
\mathcal{X}_0$, $t\in\R$, $n\in\N$, and $\Lambda_{\nu,k} \subseteq \Lambda_n$,
\begin{equation}
\label{ap4}
W_{\nu,k}\big( \Phi_t^{(n)}(x)\big) \leqslant C_2 
\big\{Q(x)\big[\log(e+n) + k^d \big]+ t^4 \big\},
\end{equation}
where $W_{\nu,k}(\cdot)$ is defined in \eqref{p3}.
\end{lemma}

\medskip\noindent
{\bf Proof.}
For notational simplicity we consider the case $t>0$. By the equations
of motion,
\begin{equation*}
\frac{d}{dt} W_{\nu,k}\big( \Phi_t^{(n)}(x)\big) = - K \: 
{\sum}^* \: \big[q^{(n)}_i(t)-q^{(n)}_j(t)\big] \, p^{(n)}_i(t),
\end{equation*}
where $\sum^*$ denotes the sum over all the sites 
$i\in\Lambda_{\nu,k}$ and $j\in \Lambda_n\setminus\Lambda_{\nu,k}$
such that $|i-j|=1$. Recalling the assumptions on the one body 
potential $U(\cdot)$, we have
\begin{eqnarray*}
\bigg| \frac{d}{dt} W_{\nu,k}\big( \Phi_t^{(n)}(x)\big) \bigg|
& \leqslant & C \, \sum_{i \in\Lambda_{\nu,k}} \big|p^{(n)}_i(t)\big|
\sum_{j:|j-i|\leqslant 1}\big| q^{(n)}_j(t)\big| 
\\ & \leqslant  & C \,
\Big[W_{\nu,k}\big( \Phi_t^{(n)}(x)\big)\Big]^{1/2}
\Big[W_{\nu,k+1}\big( \Phi_t^{(n)}(x)\big)\Big]^{1/4}.
\end{eqnarray*}
On the other hand, $W_{\nu,k+1} \leqslant \sum_{\nu'}' W_{\nu',k}$, 
where $\sum_{\nu'}'$ denotes the sum over all the sites $\nu'$ such that
$|\nu_\ell-\nu_\ell'|\leqslant 1$ for some $\ell = 1,\ldots,d$; then, 
letting
\begin{equation*}
\overline W_k(t) = \max_{\nu : \Lambda_{\nu,k} \subseteq \Lambda_n} 
W_{\nu,k}\big( \Phi_t^{(n)}(x)\big),
\end{equation*}
we get
\begin{equation*}
\bigg| \frac{d}{dt} W_{\nu,k}\big( \Phi_t^{(n)}(x)\big) \bigg|
\leqslant \overline W_k(t)^{3/4}.
\end{equation*}
Now, by integrating the last inequality from 0 to $t$ and taking the
maximum for $\nu$ such that $\Lambda_{\nu,k} \subseteq \Lambda_n$,
\begin{equation*}
\overline W_k(t) \leqslant \overline W_k(0) + C \int_0^t \!ds\, 
\overline W_k(s)^{3/4},
\end{equation*}
whence $\overline W_k(t) \leqslant \big( \overline W_k(0)^{1/4}+ C t \big)^4$.
But from the definition \eqref{p4} we have that $\overline W_k(0) \leqslant 
Q(x)\big\{2\big[\log^{1/d}(e+n)+k\big] +1\big\}^d$. The bound 
\eqref{ap4} follows from the previous estimates. 
\qed

\bigskip
We now prove the existence of the limit \eqref{ap1}. Again we 
consider the case $t>0$. We define:
\begin{eqnarray*}
\delta_i(n,t) & = & \big|q^{(n+1)}_i (t) - q^{(n)}_i (t)\big| +
\big|p^{(n+1)}_i (t) - p^{(n)}_i (t)\big|, \\
u_k(n,t) & = & \max_{i\in \Lambda_k}\, \delta_i(n,t), \qquad\quad
k\leqslant n, \\
d_n(t) & = & \max_{s \in[0,t] }\, 
\max_{i\in\Lambda_n}\, \big\{\big|q^{(n)}_i(s) - q^{(n)}_i(0)\big|
+ \big|p^{(n)}_i(s) - p^{(n)}_i(0)\big|\big\}. 
\end{eqnarray*}
From \eqref{ap2} and \eqref{ap3}, for any $i\in\Lambda_k$, we have
\begin{eqnarray*}
\delta_i(n,t) & \leqslant & \int_0^t\!ds\, 
\big|p^{(n+1)}_i (s) - p^{(n)}_i (s)\big| 
\\ && + \, \int_0^t\!ds\, 
\big| F_i(\Phi_s^{(n+1)}(x)) - F_i(\Phi_s^{(n)}(x)) \big|
\\ & \leqslant & \int_0^t\!ds\, \big[ 1 + 2d K + \big| U''(\xi_i^n(s))
\big|\big] \, u_{k+1}(n,s),
\end{eqnarray*}
where $\xi_i^n(s)$ is a point in the interval with endpoints 
$q^{(n)}_i (s)$ and $q^{(n+1)}_i (s)$. By \eqref{ap4}, for any
$s\in [0,t]$,
\begin{equation*}
\big|U''(\xi_i^n(s))\big| \leqslant C \big[1+q^{(n)}_i (s)^2 + 
q^{(n+1)}_i (s)^2\big]\leqslant C \big\{Q(x) \log(e+n) + t^4 \big\}^{1/2},
\end{equation*}
so that, setting $\varphi_n(t,x) = \big\{Q(x) \log(e+n) + t^4 \big\}$
and taking the maximum for $i\in\Lambda_k$, we arrive at the following
integral inequality,
\begin{equation*}
u_k(n,t) \leqslant C\, \varphi_n(t,x)^{1/2} \int_0^t\!ds\, u_{k+1}(n,s),
\end{equation*}
which can be solved by iteration getting, for any $n>k$,
\begin{equation*}
u_k(n,t) \leqslant C^{n-k} \, \varphi_n(t,x)^{(n-k)/2}\,
\frac{t^{n-k}}{(n-k)!} \, d_n(t).
\end{equation*}
By using the Stirling formula and observing that 
\begin{equation*}
d_n(t) \leqslant t \,  \max_{s \in[0,t] }\, \max_{i\in\Lambda_n} 
\big\{ \big|p^{(n)}_i(s) \big| + \big|F_i(\Phi_s^{(n)}(x))\big| 
\big\} \leqslant  C\, t\, \varphi_n(t,x),
\end{equation*}
we finally obtain
\begin{equation*}
u_k(n,t) \leqslant \frac{C^{n-k} \,t^{n-k+1} \, 
\varphi_n(t,x)^{(n-k+2)/2}}{(n-k)^{n-k}}.
\end{equation*}
Choosing $n_k = n_k(t,x) = 2k + C_3\big[1+Q(x)+t\big]^4$ with
$C_3$ large enough, it is easily seen that $u_k(n,t) \leqslant 2^{-(n-k)}$
for any $n\geqslant n_k$. It follows that $u_k(n,t)$ is $k$-summable, 
which implies the existence of the limit \eqref{ap1}. Indeed,
we have also an estimate on the rate of convergence: for some
constant $\delta >0$,
\begin{equation}
\label{ap5}
\big|\Phi_t(x)_i - \Phi_t^{(n_k)}(x)_i \big| \leqslant  
e^{-\delta n_k}\qquad \forall\, i\in\Lambda_k.
\end{equation}

The proof of the differentiability of $\Phi_t(x)_j$ with respect
to $x_i$ is quite standard and we omit the details. Note however that we
explicitly solved the corresponding variational equation (for $i=0$),
see Eqs.\ \eqref{p14}--\eqref{p16}.

Finally, we prove the estimate \eqref{p7}. Obviously, this also shows
that $\Phi_t(x)\in\mathcal{X}_0$ for any $t\in\R$. Given $\nu\in\Z^d$
and $k\in\N$, we set $n_* = n_{|\nu| +k}(t,x)$ and estimate 
\begin{equation}
\label{ap6}
W_{\nu,k}\big( \Phi_t(x)\big) \leqslant 
W_{\nu,k}\big( \Phi_t^{(n_*)}(x)\big)
+ \big| W_{\nu,k}\big( \Phi_t(x)\big) - 
W_{\nu,k}\big( \Phi_t^{(n_*)}(x)\big)\big|.
\end{equation}
By \eqref{ap4}, for any $k>\log^{1/d}(e+|\nu|)$,
\begin{eqnarray}
\label{ap7}
W_{\nu,k}\big( \Phi_t^{(n_*)}(x)\big) & \leqslant & C_2 
\big\{Q(x)\big[\log(e+n_*) + k^d \big]+ t^4 \big\} \nonumber
\\ & \leqslant & C \, (2k+1)^d \big\{Q(x) \log(e+Q(x)) + t^4 \big\}. 
\end{eqnarray}
On the other hand, 
\begin{eqnarray}
\label{ap8}
&& \big| W_{\nu,k}\big( \Phi_t(x)\big) - 
W_{\nu,k}\big( \Phi_t^{(n_*)}(x)\big)\big| \, \leqslant \, (2k+1)^d 
\nonumber  \\ && \qquad\qquad \times \max_{i\in\Lambda_{\nu,k}} \,
\bigg\{ \frac 12 \big|p_i (t)^2 - p^{(n_*)}_i (t)^2\big| +
\big|U(q_i (t)) - U(q^{(n_*)}_i (t))\big| \bigg\} \nonumber  \\&& 
\quad\quad \, \leqslant \, C \, (2k+1)^d \, \varphi_{n_*}(t,x)^{3/4} \, 
\max_{i\in\Lambda_{\nu,k}} \big|\Phi_t(x)_i - \Phi_t^{(n_*)}(x)_i\big|
\nonumber  \\&& \quad\quad \, \leqslant \, C \, (2k+1)^d \, \varphi_{n_*}
(t,x)^{3/4} \,e^{-\delta n_*} \, \leqslant \, C,
\end{eqnarray}
where we used \eqref{ap5}. From \eqref{ap6}, \eqref{ap7}, and 
\eqref{ap8} the bound \eqref{p7} follows. 
\qed

\end{document}